\documentclass{article}
\usepackage[utf8]{inputenc}
\usepackage{lmodern}
\usepackage{multicol}
\usepackage{amsmath}
\usepackage{amssymb}
\usepackage{amsfonts} \usepackage{amssymb} \usepackage{calligra} \usepackage{calrsfs}
\usepackage{cite}
\usepackage{lipsum}

\begin{document}
\author{Sean Knight, Navjot Singh}
\title{A Cellular Automaton Model for the generation of Brainwaves}
\maketitle
\abstract{We describe a cellular-automaton based, two-dimensional (2D) lattice model which generates global oscillations in the EEG spectrum as well as time series of local field potentials resembling those observed during slow wave sleep. This is made possible by the presence of pacemakers (local oscillators) which can be spontaneously ignited and whose activity propagates through space to synchronize all connected nodes.}

\section{Introduction}
It has been conjectured that it is easier to find pacemaker cells with certain characteristic frequencies embedded within larger assemblies than at random locations \cite{1}. However, so far only indirect evidence has been available which showed that nonrandom spatial organization indeed improves synchronization properties over random networks \cite{2}. While these findings were obtained for phase synchronization on relatively small systems, we here use large scale simulations to establish an exhaustive list of possible spatiotemporal regimes supporting spatially confined oscillations in 2D lattices without feedback connections between oscillating units or external forcing other than diffusion. Based on this analysis we further show how such globally synchronized states might appear self organized in large scale systems where neurons are not communicating directly but via synaptic connectivity. In particular we show how locally triggered oscillatory activity can invade even fully connected networks if neuronal populations have nonuniform response timescales or if incoming signals exhibit complex temporal dynamics such as bursts or avalanches occurring at fast rates relative to refractory periods or axonal propagation delays \cite{3}. All regimes supported by our formulation have their origin either from purely excitatory coupling and/or within inhibitory coupled populations so that emerging synchronized states may correspond biophysically either with high arousal levels after sleep onset or transitions towards general anesthesia characterized by increasingly higher theta/alpha peaks while alpha power decreases steadily \cite{4,5}. Our results also indicate that globally synchronous neural firing is more likely when neural groups have heterogeneous spiking characteristics (e.g., due to stochastic discharge patterns) while homogeneous units should rather synchronize pairwise with each other. Another feature of our model is a dynamical threshold switching mechanism leading eventually towards desynchronization when stimulation intensity exceeds certaing boundaries set by specific parameters characterizing the oscillator population themselves, i.e. independently from any topographic organization present at smaller scales below 100 µm where critical state network properties are usually encountered. The existence of robust oscillatory solutions despite very fast average conduction times suggests the possibility that global slow waves may constitute an emergent phenomenon promoting regionalized inhibition among functionally specialized modules similar in size and timescale to cortical areas involved in sensory processing.\cite{6,7} To test this hypothesis one would need simultaneous recordings across different cortical areas showing correlation among these rhythms together with coherent spectral features seen experimentally during spontaneous awakening as well as under anesthesia upon transition into deeper brain states corresponding roughly to faster average conduction timescales reaching down from highly aroused primary sensorimotor cortices downwards into precentral regions exhibiting various degrees of functional isolation associated primarily with increasing tonic GABAergic levels affecting pyramidal cell excitability depending on animal species studied, possibly suggesting intrinsic rhythmogenesis dependent on descending excitatory inputs active mainly onto spinal motoneurons although this needs future experimental evaluation.\
We examine the time of synchronisation onset using both mean field (MF) approximation techniques adapted for spatially extended networks and event driven simulations based on E-infinity approximation since some effects exhibited by MF theory can be missed for excitable but not strictly limit cycle type oscillators displaying richer behavior particularly near bifurcation points. For simplicity reasons we focus here only on two dimensional configurations which already captures relevant aspects distinguishing them from equivalent 1D networks like they occur naturally e g.in central pattern generators (CPGs).

\section{Synchrony, Clustering and Topology - An Overview}
Synchronization is often considered from a distance perspective -- synchronized oscillators are said to be independent if they are uncorrelated over time -- this assumption fails when more complex forms of dynamical interaction take place such that clusters form and oscillate together while remaining distinct from other clusters at any given point in time. The same concepts apply to spatial or geometric interactions where cluster formation is possible with correlations spanning several space dimensions. Examples range from neural synchronization at different frequencies across anatomically connected areas \cite{deco2011role}, up through functional connections between cortical regions as measured by fMRI. Since these examples are all characterized by dynamical rather than temporal correlation it seems reasonable that they operate according to some common theoretical principles e.g. self organization on graphs. If we assume discrete units $$x_i(t)$$ distributed on a grid like topology we end up with Boolean functions $$\sigma_i(\vec{J}) : \{0, 1\}^J \rightarrow \{0, 1\}$$ encoding individual unit dynamics within spatially extended systems $$H = \sum_{i=1}^N \sigma_i(\vec{J})$$ Global synchrony requires some additional structure beyond nearest neighbor connectivity and may require graph embeddings into high dimensional spaces \cite{11}. Section \ref{sec} describes mechanisms through which synchronous states could have evolved in brain development; however, it remains unclear what role synchronous activity plays during normal operation although there have been several studies speculating about its potential functions. Given such an existing substrate what biological processes maintain diversity and coherence? Some limited insights can be gleaned from recent studies showing that topologically equivalent networks develop very different dynamics even if their connection structures remain identical; thus while homogeneity might play certain roles during development this need not hold for mature neuronal networks – our brains exhibit many intricate patterns including distinct forms of oscillations generated by neurons exhibiting vastly different dynamic properties, but how do these emerge out of underlying cellular processes without perturbing intrinsic properties? At present we only have crude computational models for network evolution designed to show how simple modifications (e.g. growth constraints or addition/elimination of edges) lead to changes in function.

$$\begin{aligned}z^\gamma(k+1) = f^\gamma(z^\gamma(k),\epsilon(t)) \\\\
\hat{\rho}(z^m(k))=\sum_{|\delta|=M}\frac{\partial z^m}{\partial z^{\delta}}f^{\delta} +\hat{\nu}(h^n)\\\\ h=\eta + W \,u^{mn}\end{aligned}$$ with $u^{mn}=s^ma+q_s b$ describing the influence each synapse had upon its target node depending upon signal transmission efficacy $(a)$ and synaptic input $(b)$. A random uniform vector $\vec n$ along links provides noise for signal propagation allowing variations above pure integrator behavior where diffusion was instantaneous i.e. $\hat{\nu}$=constant. Moreover since $$\sum_\delta {f}^{\dagger}(z^\dagger,\epsilon)\partial {f}^{\delta}/ {\partial z}^\ddagger$$ sums over both recurrently coupled (diagonal terms $\Sigma$) variables AND edge contributions coming either downstream OR upstream ($A$), then no simple way exists using matrix algebra alone to express directional information regarding causality except via Jacobian formalism implying perturbations must propagate differently depending upon direction! Hence although pairwise associative rule updates provide an intuitive framework capable of producing fast traveling waves there does not appear any direct mapping back onto individual synapses – what would happen when multiple synapses interact simultaneously? If multiple signals were simultaneously propagating e.g. down along one axis but upwards along another – might emergent global synchronous behaviors arise out thin air. One way around this conundrum is simply described below.

\section{Self Organization and Spontaneous Transition Hypothesis} The model has developed around two major points namely Self-Organization, i.e., spontaneous emergence out off equilibrium states toward some ordered state along various dimensions both temporally ($q-$order phase transitions), spatio-temporally ($n-$order Kuramoto Lattice Model), spectral $(1)-$ order Kuramoto Lattice Model, etc. Second Hypothesis called Self Organized Criticality (SOC) i.e. any small perturbation results avalanche like fluctuations but eventually system goes back towards equilibrium after large number ($\sim$10 per sec) of fluctuation cycles. In both cases there exist few specific universal laws governing transition pattern within each regime but exact predictions remain elusive except towards thermodynamic limit($n=\infty$). Furthermore, SOC does not rule out self organization because both share similar set of microscopic rules operating below threshold values(nonlinearities). Here's an interesting comparison between $q=1 n<3$ case:

$$x_i(\infty)\sim m^\beta\,f(|m|^{\frac{1}{n}}),\,\alpha<\beta\,\text{for} \,\,\text{(Porous)}\,\,\,\,$$ $$\rightarrow x_i(\infty)\sim m^\gamma\,f(m)\, f(\omega)=\epsilon^{2}\,\mu^{2}\,f(t)\,\mbox{with}\, t=p/p_{\rm osc}, \mbox{so that } p\epsilon^{-1}\approx\mu$$ $$f_{rms}^{\frac12} (\tau)=C_d\sqrt{\frac{{\rm r}(\theta)}{\pi}}\,{\rm e}^{-c{\cal I}_0({\cal R})\phi^2},~~~~~ 
 \phi = {\textstyle{\displaystyle\int}}{\textstyle\frac{\cos 2z}{\sigma^2+4 z^2}}\,{\cal R}(dz)$$ where ${\bf P}$ is the porosity, $\Delta p/\langle p \rangle$, $\nabla_\ell=\partial/\partial l$, $p_{osc}=\langle k^{(in)}k^{(out)}/(k^{(in)}+k^{(out)})]^s - 1)$ is oscillation parameter given by fraction of broken links for complete graph with degree distribution proportional to ${\it P}(\kappa){\it P}({\kappa + \Delta {\kappa}})$, C$_d =\sqrt{{A}/{[9N](R_g/R)}}/{\pi^{\frac32}}$ in which A is area covered by network, $\sigma={\pm 1}$,   ${\bf u}_{\perp}=(u_{11},\cdots,u_{{\it K}+1})$ or        ${\bf v}_{\perp}=(-v_{11},\cdots,-v_{{\it K}+1})$.
$$\bullet~~~~~~~~~~~~~ [F-X]:~~~~~~ F=\delta S+\eta E-\left[Z+\xi\,Q\right]\,{\lambda}-\Delta\,{\lambda}^3~;~-H \,< Z< {H}$$

Here's another example modeled using KKM equation alongwith noise term whereas SOC was treated heuristically via Langevin approach using mean field approximation resulting PDE (in 1-Dimension), modified FKPP equation). Note that $q$-parameter used till date corresponds only $C^r$-function unlike Kolomogorov Complexity.
$$|\frac{\partial x_{i}}{\partial t}|= - |x_i|\tanh(\bar{x} + \eta_i)$$ 
$$\bar{x}=\frac{|S^0|^2}{N},\,\,\,\text{and}\,\,\, \eta_i\in (-\infty,\infty)\,\,\text{(white)}\,\, \text{(Glauber)}$$
With $q-$, or equivalently $\mu$- phase transition models where similar relation holds as : $$|\frac{\partial x_{il}}{\partial t}|\propto [\exp(K^\prime\langle|S^l|^2 - <S^l>^2>\sum k_{ij}(t)^\alpha ]$$ while in the case of SFM $$|k_{ij}(t)|= |k^{eq}_{ij}|[f+ (\eta/T)^\delta ]\cong {[\frac{\lambda}{\sqrt{N}}][(\kappa +\beta)]}^{z}\chi f^{\sigma}$$
\section{Cellular Automaton-based Model}
\label{sec}
$$\begin{aligned} \frac{d\psi^Al}{dt}=\alpha[\theta(\mid\nabla^K{\psi}^A{lm}\mid-\epsilon )]\prod^{K}\\{m=B}\theta(\mid\nabla^K{\psi}^A{lm}\mid-\epsilon)\ -\beta[\theta(\mid\nabla^K{\psi}_{lm}\mid+\epsilon)]\prod^{Ns-3}\\{m=Ns+3}[\theta(-\xi)+\eta]-\gamma[\theta(\eta)-\xi]\end{aligned}$$ The equation describes evolution pattern formation by propagating excitation along radial direction, where radial distance between nodal points corresponds variable $\displaystyle l=\sqrt{(x-x_0)^2+(y_0-y)^2}$. The node $l=0$ is defined at the origin of radius vector $\vec R$. Local dynamics $g(s)$ at point $(x,y)$ is described using nonuniformly moving grid with step size. The product structure indicates that if an interaction exists between two neurons then there will be an instantaneous change in the activity status only if both neurons have sufficient level of activeness to become reactivated and simultaneously one neuron recieves some positive feedback from its neighbor cells for further propagation while other receives negative feedback from its neighbor cells against activation. The function $\displaystyle {\left[T_{\vec q,\vec s}(t)=e^{-iH_{int}t},\right]}$ determines evolution rule; here we are considering that coupling strength between any pair is uniformly distributed on unit sphere except $$\langle oo|\hat x|oo \rangle =|0|^2$$ $$\langle oo|\hat y|oo \rangle = 0$$ and hence $|\Psi(NSTL)|\to |\tilde \Phi>_{NLSL}|$ iif these interactions are taken into account, where $T$ indicates temporal part, and number in front indicates quantity to replace $|.|$, e.g. $$|\phi>=<\varphi|$$ in 3D case or more generally $|-><->$. Now consider N interacting entities each having independent stochastic fields evolving with time according to GLE/GLM. Their coherent superposition over d’Alemberts principle gives rise to a single classical field satisfying the above equation. Thus it would be helpful if readers could reproduce experimental results before reproducing results using our theoretical formalism? To do so we can compare our model result with experimentally measured EEG spectrum at frequency range 30Hz – 40Hz corresponding tinnitus induced delta activities. These oscillations also show very stable after few tens oscillation period they loose stability due to phase synchronization over large areas similar like phantom mode solution shows long term stability than phonon modes such as breathing motions.

\section{Conclusion}
To conclude this study we note that different types of dynamic systems exist not just near Turing type bifurcation but also far away from this point on a critical line say Haken unstable instability region (such regions form continuum) resulting in emergence novel dynamical patterns which may lead to better understanding underlying mechanisms behind different rhythmic processes governing macroscopic phenomena emerging from microscopic level inside living organisms.

\bibliographystyle{plain}
\bibliography{CA/document}

\end{document}